\documentclass[aps,prb,superscriptaddress,twocolumn,showpacs,amsmath,amssymb,showkeys]{revtex4-1}

\usepackage{graphicx}
\usepackage{epstopdf}
\usepackage{xcolor}
\usepackage{amsmath}
\usepackage{gensymb}

\def\nm{\,\textrm{nm}}
\def\keV{\,\textrm{keV}}
\def\nA{\,\textrm{nA}}
\def\pA{\,\textrm{pA}}
\def\fps{\,\textrm{fps}}
\def\kV{\,\textrm{kV}}
\def\mm{\,\textrm{mm}}
\def\mrad{\,\textrm{mrad}}
\def\tAA{\,\textrm{\AA}}
\DeclareUnicodeCharacter{2212}{-}

\begin{document}

\title{St4DeM \\
A software suite for multi-modal
4D-STEM acquisition techniques}

\author{\bigskip Toni Uusimäki}
\email[Email: ]{tonimuusimaki@gmail.com}
\affiliation{Department of Materials and Environmental Chemistry, Stockholm University, Stockholm 106 91, Sweden}

\author{Cheuk-Wai Tai}
\affiliation{Department of Materials and Environmental Chemistry, Stockholm University, Stockholm 106 91, Sweden}

\author{Tom Willhammar}
\affiliation{Department of Materials and Environmental Chemistry, Stockholm University, Stockholm 106 91, Sweden}

\author{Thomas Thersleff}
\affiliation{Department of Materials and Environmental Chemistry, Stockholm University, Stockholm 106 91, Sweden}

\author{Hasan Ali}
\affiliation{Department of Materials Science and Engineering, Uppsala University, Uppsala 751 03, Sweden}

\author{Seda Ulusoy}
\affiliation{Department of Materials Science and Engineering, Uppsala University, Uppsala 751 03, Sweden}

\author{Meltem Sezen}
\affiliation{Nanotechnology Research and Application Center, SUNUM, Sabanci University, Istanbul 34956, Turkey }

\author{Bora Derin}
\affiliation{Department of Metallurgical and Materials Engineering, Istanbul Technical University, 34469 Istanbul, Turkey}

\begin{abstract}
\bigskip
A suite of acquisition applications related to the 4D-STEM technique is presented as a software package written within the Digital Micrograph environment, which is a widely used software platform in worldwide electron microscopy laboratories. The 4D-STEM technique allows the acquisition of diffraction patterns at each electron probe position in a scanning transmission electron microscope map. This suite includes 4D-STEM acquisition, ptychography, EELS/EDS spectrum imaging, tomography and basic virtual visualization and alignment methods on 4D data including incoherent differential phase contrast analysis. By integrating electron tomography with 4D-STEM and EELS SI, St4DeM enables the acquisition and analysis of 7-dimensional data.

\end{abstract}

\date{\today}
\maketitle

\section{Introduction}

The unequivocal imaging resolution provided by the aberration-corrected scanning transmission electron microscope (STEM) \cite{Pennycook2011}, and its ability to study nanoscale phenomena locally by converging the electron beam onto the sample at subatomic probe sizes \cite{Krivanek2015} have made it an unmatched technique in the field of material science \cite{Wu2018}. Furthermore, accompanied by the computing power and memory of modern computers and fast direct electron counting cameras \cite{Levin2021}, STEM can be used not just to acquire a one-dimensional intensity but simultaneously to record a 2D converged beam electron diffraction (CBED) image at every probe position within the scan.

Traditionally in STEM, a convergent beam of electrons is focused onto the sample, and a 2D raster image is recorded. The intensity distribution created by the interactions of the electrons and the sample can be acquired in various ways including bright field (BF), dark field (DF) or high angle annular DF (HAADF), where a single value of electron counts is contributed to a single pixel. In 4D-STEM \cite{Ophus2019}, on every pixel of a 2D STEM image, a CBED or diffraction pattern (DP) is acquired using a charge-coupled device (CCD) or pixelated camera – hence giving additional 2D angular diffraction information. Coupling the speed of modern cameras with hardware synchronized beam control yields staggering acquisition speeds – as $100k$ frames per seconds ($\fps$) are being reported \cite{Chatterjee2021}. Rather than a specialized single technique, 4D-STEM is a family of different acquisition modes, where either a real or diffraction space image is acquired at every pixel. It has been successfully used in visualizing electric and magnetic fields \cite{Fang2019,Nguyen2016}, strain \cite{Allen2021} and orientation mapping \cite{Jeong2021}, structure determination \cite{Shukla2016} and ptychography \cite{Muller2019}. It is hence likely, that the near future roadmap within the field of STEM accelerates to combine and fuse multi-modal data \cite{Thersleff2020,Thersleff2023} e.g. using 4D-STEM with electron energy loss spectroscopy (EELS) and/or X-ray energy dispersive spectroscopy (EDS); while simultaneously aspire towards multiple dimensions e.g. by applying electron tomography \cite{Weyland2004} and/or in-situ electron microscopy \cite{Zheng2017}.

\begin{figure*}[ht]
	\centering
	  \includegraphics[width=1.0\linewidth]{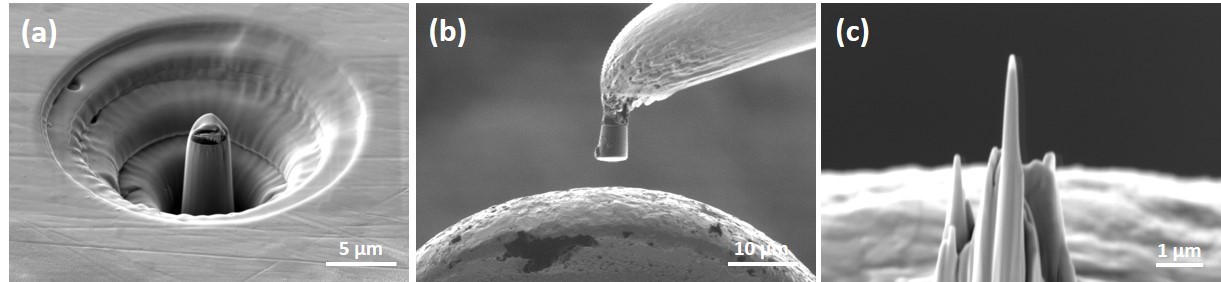}
	\caption{The preparation of needle-shaped sample using FIB processing: (a) The milling of the conic sample using circular patterns at relatively high ion currents ($1-10 \nA$); (b) Lifting out the section from the main sample using the micromanipulator and mounting the cone on top of the TEM grid using carbon deposition; (c) Fine-milling of the sample with low ion currents ($50-500 \pA$) until a needle with a dimeter less than $)100 \nm$)is reached.}
\end{figure*}

The St4DeM was written as a suite to collect and implement various 4D-STEM acquisition modes into one package, additionally incorporating energy filtering TEM (EFTEM), 3D EELS and EDS chemical mapping techniques. The software suite was written within the Digital Micrograph (DM) scripting environment \cite{Mitchell2005}, which is a popular software platform installed in most electron microscope controlling computers worldwide. Here, the basic 4D-STEM acquisition mode, 4D-STEM tomography, 4D visualization, and analysis methods are introduced, with a focus on the proof of principle of the software rather than the detailed analysis of the samples. In addition, but not shown here, AutoEM particle sizing software AutoEM \cite{Uusimäki2019} was incorporated into St4DeM, allowing 4D-STEM and EELS/EDS spectrum imaging (SI) acquisition of unlimited number of nanoparticles with serial montage imaging as used in particle sizing methods. The basic 4D-STEM imaging of St4DeM has already been explored and published in \cite{Bello-Jurado2022, Wilhammar2021}. Momentum resolved EELS was shown, as electron magnetic circular dichroism (EMCD) in \cite{Ali2023}. The software suite, the acquired original data and the analysis code used in this paper can be downloaded from \url{https://github.com/AutoEM/St4DeM} and
\url{https://zenodo.org/record/7529702#.Y8_YEHbMJP}. 

\section{Materials and Methods}

The Fe$_3$O$_4$ nanoparticles were synthesized by decomposition of Fe(acac)$_3$ in the presence of oleic acid (OA) and Na-oleate precursors, as reported in \cite{Ulusoy2024}. 6 mmol of Fe(acac)$_3$ was mixed with 12 mmol OA and 0.1 mmol of Na-oleate by adding 40 mL dibenzyl ether, 40 mL octadecene and 12 mL tetradecene solvents. The mixture was magnetically stirred and degassed at room temperature for 1h. Then, the slurry was heated up to reflux temperature ($\sim$290 \degree C) at a rate of 20 \degree C/min under Argon atmosphere and kept at this temperature for 30 min. After cooling to room temperature, the resultant product was washed 3-4 times with a mixture of toluene and ethanol (1:4) and centrifuged 6000 rpm for 5 min discarding the supernatant to remove residual organics. This step is repeated by sequential operations of dispersing in toluene, precipitation by adding ethanol and centrifuge until organic amount is less than 2 \%. Fe$_3$O$_4$ nanoparticles with cubic shape and edge length of $l = 73 \pm 9 \nm $ were synthesized.  

The carbon nanotube powder and rare earth element (REE) nanoparticles (allanite, monazite and parisite)  were dispersed in hexane and deposited onto TEM grid by drop on grid technique.

A lab-fabricated TiNi shape memory alloy (SMA) for dental applications, was prepared using focused ion beam milling at the JEOL 4601F MultiBeam System. The sample was circularly ion-milled using the annular patterns (Figure 1) at $30 \keV$ ion energy and with the ion currents ranging from $10 \nA$ down to $50 \pA$ for coarse/fine thinning and polishing stages, respectively, until an electron-transparent pin is formed. 

The software was tested and developed using a Thermo Fischer $300 \kV$ Cs-corrected Themis and a $200 \kV$ JEOL JEM-2100F microscope (Stockholm and Trondheim), with DM versions 3 and 2 respectively.  The successfully tested cameras in Themis were a local Gatan Quantum 965 ER and a remote Gatan Oneview, and on JEOL Gatan Ultrascan 1000, GIF Tridiem and Orius. The Oneview camera control software was located on an external PC, hence a boost library \cite{Karlsson2005} UDP server was created on the microscope PC controlling the beam scanning, and the Oneview image acquisition was synchronized as a client. The Oneview camera can operate at $300 \fps$ acquisition speed, however since a server/client connection was used to synchronize the acquisition and scanning, the speed was limited to $100 \fps$.

\begin{figure*}[ht]
	\centering
	  \includegraphics[width=0.8\linewidth]{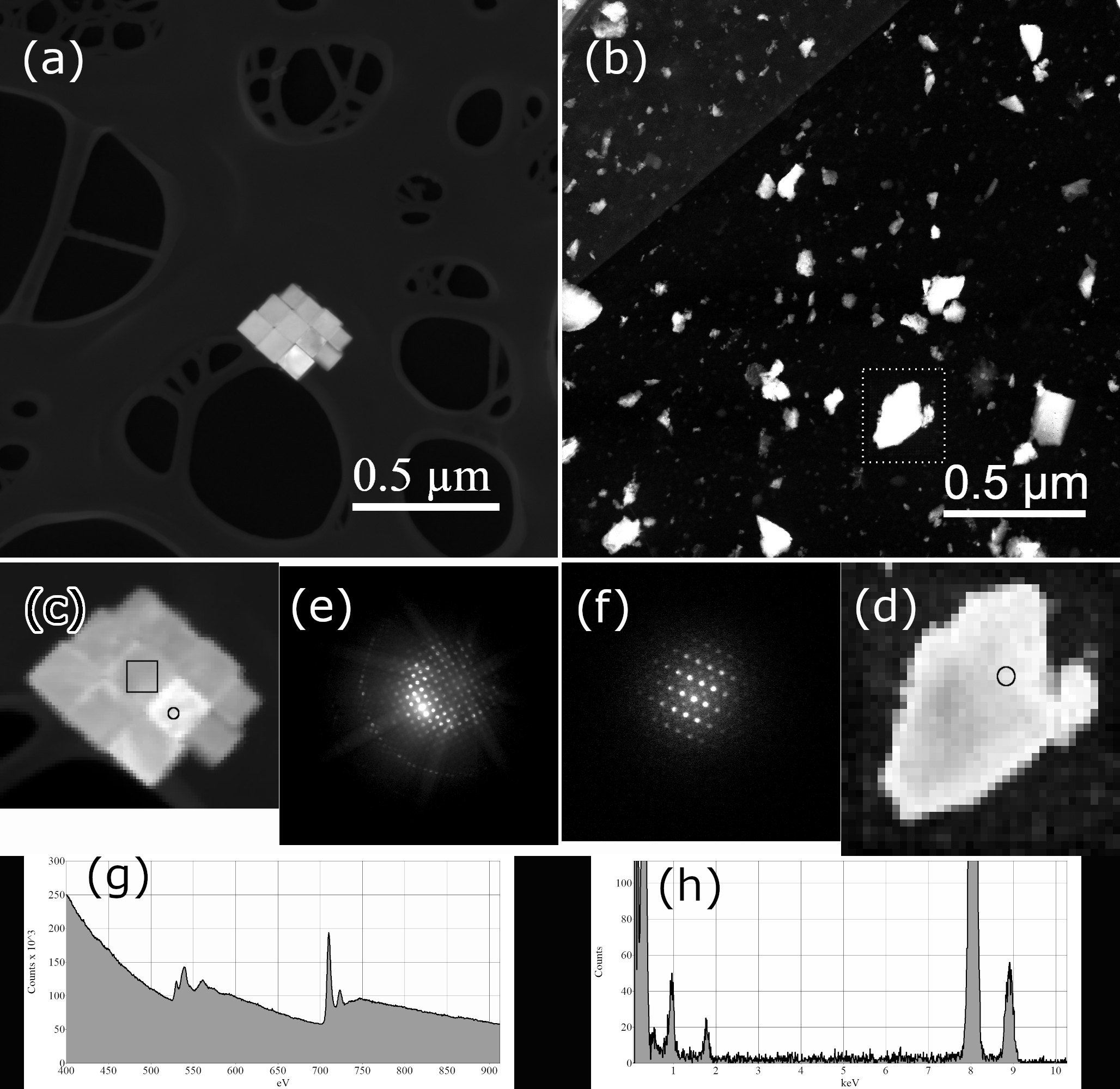}
	\caption{STEM HAADF images (a) and (b) acquired using a Thermo Fischer Themis $300 \kV$ and JEOL JEM-2100F respectively, from where the 4D-STEM images (c) and (d) were recorded. From a single pixel inside the small circles, diffraction patterns (e) and (f) are shown. Additionally, an integrated high loss EELS spectra (g) were summed from the acquired 3D SI, inside the square box in (c). Figure (h) shows the integrated EDS spectra from the whole area shown in (d).}
\end{figure*}

The images of the Fe$_3$O$_4$ nanocubes, carbon nanotubes and TiNi were acquired with Themis operated in microbeam STEM mode and using Oneview camera. Condenser aperture of $50 \mu m $, spot size 7, indicated camera length (CL) of $145 \mm$. A measured screen current of $0.046 \nA$ and indicated convergence angle (CA) of $0.21 \mrad$, were used for the first two samples, and for TiNi a CA of $0.49 \mrad$. The implemented auto-focus routine acquires fast images on either user specified region of interest (ROI) or automatically, and uses a Nelder-Mead simplex method \cite{Rudnaya2011} and variance as a sharpness function to optimize defocus. Alternatively the software can find the optimal ROI for the auto-focus routine by brute search using again the variance divided by the mean value of the rolling ROI across the image.  However, the auto-focus routine was not used in the 4D-STEM tilt series acquisition, since Thermo Fischer TEM Scripting SDK does not allow to change the defocus while it is controlled by the C3 lens, which is the case in microbeam mode. CA was minimized using the built-in Free Lens module in the Tecnai User Interface. The images of REE nanoparticles were acquired with JEM-2100F, and the CA was minimized using the free lens control as described in \cite{Plana2018}. A condenser aperture of 50 µm, indicated probe size of $1 \nm$, and a CL of “HAADF5” were used. The magnification and CL calibrations were performed for a range of magnifications and CLs with a gold-plated cross grating sample. A CCD image was acquired using a large defocus for rotation calibration between the diffraction and real space. Additionally, a defocus series was acquired with a CCD camera to get an image of the probe for a wide range of defocuse's (procedure included in the St4DeM UI).

\section{Results}

The figure 2(a) shows a normal STEM image of the Fe$_3$O$_4$ nanocubes on holey C-grid before the 4D-STEM acquisition, taken with a Themis $300 \kV$. Placing a ROI around the group of nanoparticles one can then simply acquire a 4D-STEM dataset $(86\times76\times512\times512)$ as shown in figure 2(c). Each of the pixels then corresponds to a 2D DP as depicted in figure 2(e), taken from the center of the small circle. After the acquisition one can also acquire an EELS or EDS SI within the St4DeM UI, using the same ROI giving identical spatial resolution and sampling. An integrated EELS high loss signal of O K- and Fe L-edge is shown in figure 2(g) taken from the square annotation within the middle particle (low loss signal not shown). Figure 2(b) in turn shows a STEM image of REE nanoparticles on a C-grid, taken with a JEOL 2100F microscope. Similarly, a 4D-STEM dataset $(38\times40\times512\times512)$ is presented in 2(d), and the associated DP in 2(f), taken from a single pixel inside the small circle. An integrated EDS spectrum is shown in 2(h) summed over the whole SI. Currently a commercial SI plug-in has to be present in DM for acquiring EDS/EELS SI, but is not necessary for momentum resolved 2D EELS.

\begin{figure*}[ht]
	\centering
	  \includegraphics[width=1.0\linewidth]{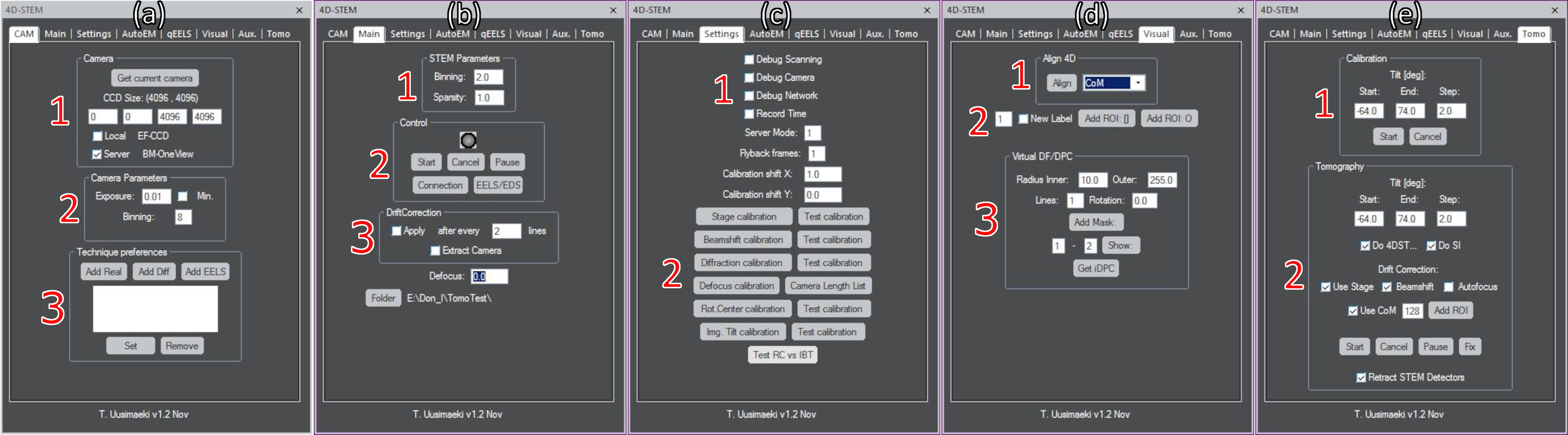}
	\caption{The St4DeM user interface within DM 3, showing (a) the Camera, (b) Main, (c) Settings, (d) Visualization and (e) Tomography tabs.}
\end{figure*}

For performance and speed, the acquisition, server\slash client configuration, scanning and visualization threads were written with the DM SDK C\texttt{++} language with Visual Studio 2008 and 2015 for DM version $2.x$ and $3.x$ respectively, and compiled as a library to be copied to the DM plugins folder. The microscope control C\texttt{++} wrapper functions were as well compiled into libraries both for Thermo Fischer and JEOL microscopes. The St4DeM UI was written with the Digital Micrograph scripting language. In figure 3 the 5 different tabs of the UI is shown. Figure 3(a) shows the camera tab, where the user can choose a local or remote camera (1), set exposure time and camera binning (2), as well as to save the current camera length, beam shift and diffraction shift settings for later and easy usage when shifting between diffraction, imaging or EELS conditions (3). A TCP client/server is used between the local and remote PC for basic communication, if the camera is not a local one; however, during acquisition an UDP or WinSock connection was tested for faster operation. Figure 2(b) shows the main tab, where the user can set STEM binning and sparsity value (1) for low-dose measurements. For example, a value of 0.5 would choose randomly (Bernoulli distribution) only 50\% of the probe positions to acquire an image. The control box (2) for starting both 4D-STEM and EELS/EDS SI acquisition as well as settings for drift correction (3), folder for auto-save and defocus value for ptychography.  Figure 3(c) shows the settings tab for debugging (1), and calibration procedures for stage, beam and diffraction shift (2). Figure 3(d) depicts the 4D alignment (1) options, visualization (2) and integrated phase contrast imaging (iDPC) options (3).  The tomography tab (Figure 3(e)) provides methods to record a traditional STEM tilt series, as well as combined with 4D-STEM and/or EELS/EDS SI. The tomogaphy stage drift calibration can be performed (1) for faster performance and the tomography box provides options for the drift corrections, autofocus and final Center of Mass -alignment (CoM) for the 4D-STEM acquisition.

\subsection{Visualization and iDPC}

The 4D data sets contain a wealth of information. The St4DeM UI provides easy real-time methods to visualize the data within DM. Figure 4(a) shows a 4D-STEM data set $(53\times51\times512\times512)$ from a Fe$_3$O$_4$ nanocube of $77 \nm$ in size. By the common nomenclature the first two axes $(53\times51)$ are generally called navigation axes, whereas the DPs $(512\times512)$ are called signal axes. By adding ROIs using the UI, one can visualize several DPs at the same time either by single ROI or summing several ROIs together. The DP from figure 4(b) is a summed image from the two large ROIs entitled with label 1 in figure 4(a), where as figure 4(c) is from a single pixel ROI with label 2. 

\begin{figure*}[ht]
	\centering
	  \includegraphics[width=0.65\linewidth]{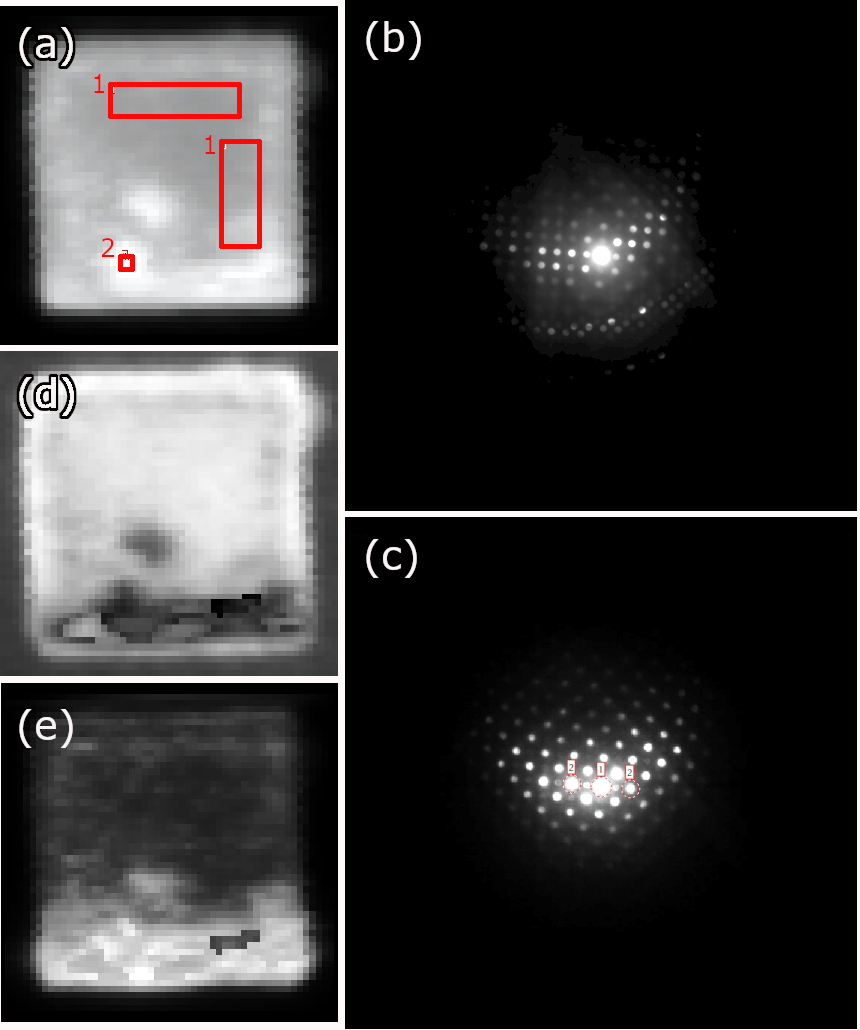}
	\caption{A 4D-STEM image of Fe$_3$O$_4$ nanocube, from where an integrated diffraction pattern (b) of two large ROIs has been extracted with label 1. Figure (c) shows a diffraction pattern from a single pixel with label 2. Figure (d) is a virtual bright field image of (a) taken from the direct beam circular label 1. Similarly figure (e) is a virtual dark field image, summed using $(-g, g)$ peaks with circular labels 2.}
\end{figure*}

Similarly, by adding circular ROIs into the DPs, one can create virtual BF, DF or HAADF images of the nanocube. Figure 4(d) shows the virtual BF image of figure 4(a), by summing all the pixels contained within the ROI labelled 1 – situated in the $(000)$ peak in figure 4(c), for every pixel on the navigation axes. Whereas figure 4(e) is a virtual DF image by summing all the pixels within the two circular ROIs labelled 2, situated on the $-g$ and $g$ spots. Annular segmented virtual images are also supported.

The St4DeM UI provides different methods to align the DPs. For low CL data sets a CoM alignment is provided, and for high CL DPs a novel method was written, which takes the autocorrelation of the DP, which is always centered, and then cross-correlates the DP to the autocorrelated image. To preserve the iDPC information, also a mean DP CoM alignment is provided. Figure 5(a) shows a 4D data set of a carbon nanotube sample with added noise, which shows very little contrast. In order to obtain iDPC information, a CoMx and CoMy images has to be calculated. If a circular ROI is present in the DP image, it is used as mask for the shift calculations as seen in 5(a) inlet image. Simultaneously the pixels per milliradians calibration and the rotation induced by projection lenses are obtained. Figure 5(b) shows the iDPC ($E_{mag}$) as magnitude of the x and y shifts and 5(c) ($E_{dir}$) depicts the direction of the shifts with associated color wheel. The iDPC methods are implemented similarly as given in \cite{GetDPC}, wherein Figure 5 (d) and 5(e) are the Charge density and potential respectively. The software automatically subtracts a fitted 2D plane from the CoMx and CoMy images to remove the d-scanning effects. One can additionally extract the x and y shift images from the 4D-STEM data using virtual annular segmented detector. 

\begin{figure*}[ht]
	\centering
	  \includegraphics[width=0.75\linewidth]{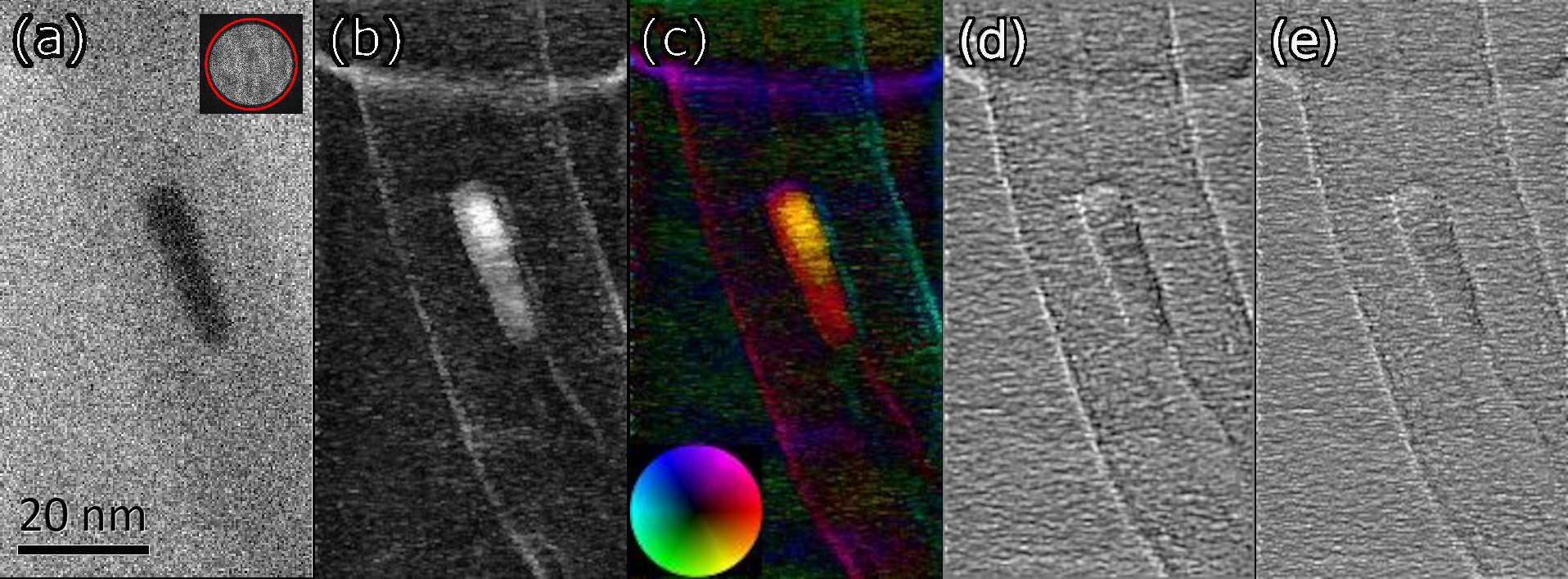}
	\caption{A 4D-STEM image of carbon nanotube with added noise (a), where the inlet DP shown in upper right corner with a red circle used as a mask. The iDPC (b) shows clear phase retrieval from the noisy input image. The image (c) shows the phase components with color wheel. In (d) and (e) are the charge density and potential respectively.}
\end{figure*}

\bigskip
\subsection{7D-STEM}

Typical STEM signals are based on transmission characteristics, and it is hence difficult to access the physical dimension of the specimen parallel to the electron optical axis. A solution to this common problem is to apply electron tomography, where the specimen is tilted using the goniometer and an image is recorded on every tilt step compromising as a tilt series, which can then be reconstructed using the methods generally described in the field of inverse problems. St4DeM allows automatic tilt series acquisition combined with 4D-STEM and/or EELS/EDS SI. There exists numerous sophisticated reconstruction software to back-calculate the object from the tilt series typically using back-projection methods \cite{Radermacher2007}. Reconstructions are calculated one plane at a time, thus reducing it to a 2D problem. For every tilt angle a horizontal 1D line profile is taken from a single image from the tilt series, multiplying its Fourier transform (FT) with some weighting function (Hanning, Hamming filter etc..) to amplify the high frequency content – and then back-projecting the inverse FT´s intensities into the 2D digital reconstruction at an angle corresponding to the goniometer tilt. These intensities do not exactly match the regular grid of the 2D reconstruction image, hence linear or cubic interpolation schemes are necessary for high fidelity reconstructions.

The 4D-STEM tomographic reconstruction from a 4D-STEM tilt series depicts a similar path as above, yet some discrepancies arise. If first, by considering a single voxel inside the reconstruction, the accumulation of DPs into the 3D DP by their tilt angle is very similar to 3D electron diffraction (3D ED) \cite{Zou2021}, where a near parallel TEM beam is used to illuminate the whole particle at once. The same principle applies to 6D reconstruction as well, except now we accumulate the DPs along the line integral of the $3D_{real}$ reconstruction to every $3D_{diff}$ voxel within that line by their tilt angle.

Rather than accumulating single value intensities by the back-projected line profiles with every tilt angle, the 2D DPs are projected along the ray by reconstructing a 3D DP to every voxel along the ray – by rotating the 2D DP by its tilt angle.  The final reconstructed object has the dimensions of  $[X_r,Y_r, Z_r][X_d, Y_d, Z_d][E]$, where $[X_r,Y_r, Z_r]$ are the navigation axes and $[X_d, Y_d, Z_d]$ and $[E]$ are the signal axes for diffraction and energy space respectively. Since a 2D DP is projected along the ray into the reconstruction, rather than one single intensity value - no weighting nor Fourier transforms are used. Also, interpolation between two adjacent DPs seems similar as to purposely introducing problematic dynamical diffraction behavior to already there which is present, since the 4D-STEM tilt series in this study was acquired without using any precession electron diffraction hardware. Hence for this research, only nearest neighbor interpolation was used. The energy dimension can be populated one energy step at a time using regular single value reconstruction techniques.

\begin{figure*}[ht]
	\centering
	  \includegraphics[width=0.7\linewidth]{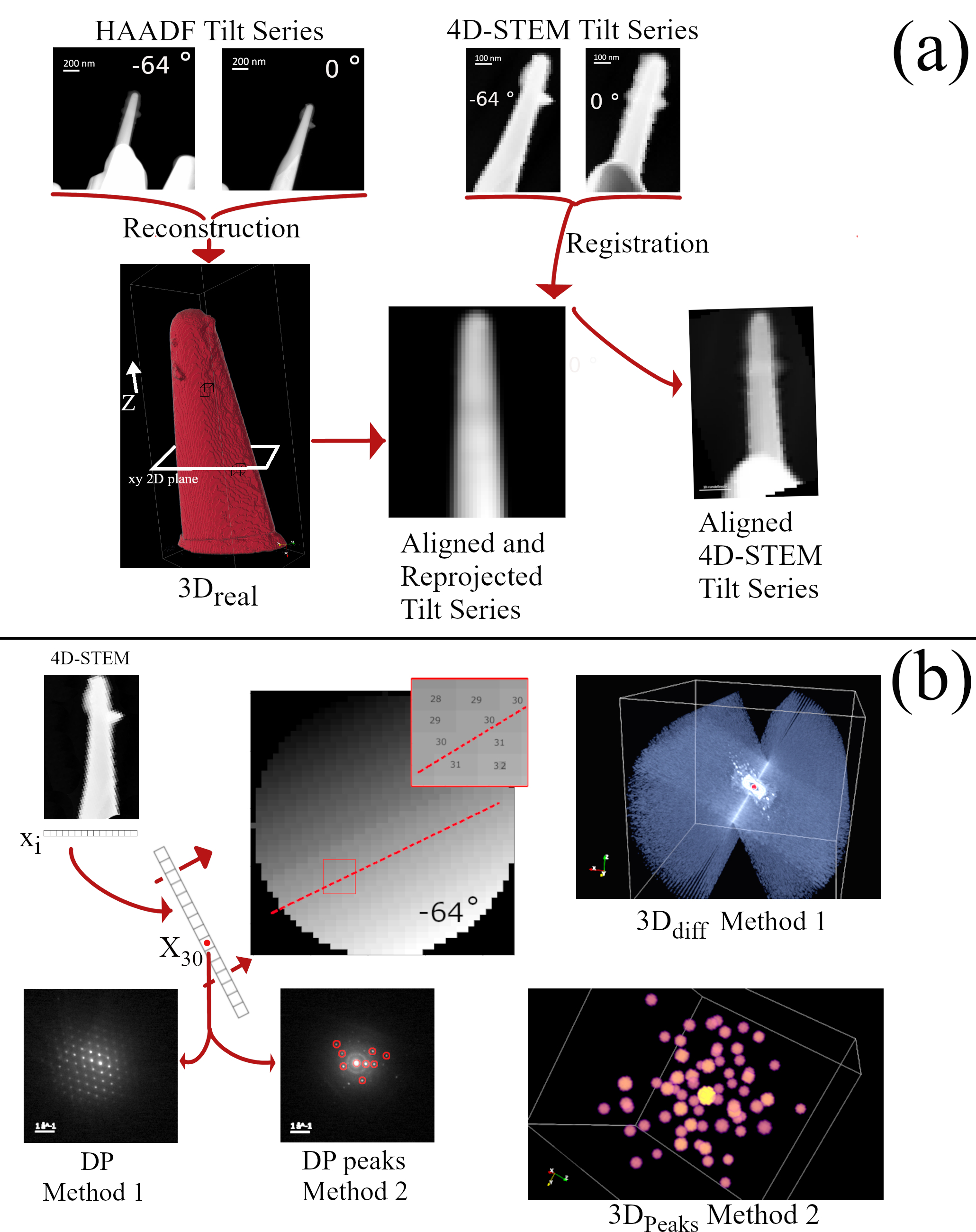}
	\caption{6D-STEM workflow. (a) STEM HAADF tilt series is reconstructed into $3D_{real}$ using SIRT. The reconstruction is then resampled and re-projected as a new tilt series to which the 4D-STEM tilt series is registered. (b) The 2D DPs are projected into $3D_{dreal}$ using the inverse Radon map as a guide, such that e.g. in method 1, the DP at the 4D-STEM x-axis index 30 is inserted on every voxel on the projected (red) ray, and rotated with the corresponding goniometer tilt angle. Hence every voxel of $3D_{dreal}$ contains a unique $3D_{diff}$. Similarly in method 2, the found 2D peaks on every x index are projected along the ray while the goniometer tilt gives the 3D coordinates of the peaks.}
\end{figure*}

The workflow of the 6D-STEM was based on the single value STEM HAADF tilt series recorded from -64\degree$\:$to 74\degree$\:$tilt angles, and was cropped, aligned and processed using the TomViz software \cite{Schwartz2022}. The 3D reconstruction ($3D_{real}$) was calculated using simultaneous iterative reconstruction technique (SIRT) with 10 iterations. The $3D_{real}$ was then down resampled ($3D_{dreal}$) to match the sampling of the 4D-STEM images and re-projected to form a new virtual tilt series, to which the 4D-STEM tilt series were registered using SimpleITK procedures and Similarity2DTransform \cite{Yaniv2018}. Using this approach, it is given, that the 4D-STEM tilt series is aligned properly with respect to the $3D_{dreal}$ reconstruction. These and the following procedures were made using Jupyter Notebook \cite{Kluyver2016} with the aid of Hyperspy \cite{francisco_de_la_pena_2015_27735} and Pyxem \cite{Cautaerts2022} libraries. From here two different options are available: full $3D_{dreal}$-$3D_{diff}$ reconstruction (Method 1) or $3D_{dreal}$-$3D_{peaks}$ list object (Method 2). Former consist of a real 3D reconstruction, where every voxel contains yet another 3D DP, therefore consisting of 6 dimensions. The latter can be constructed after using a normal peak search algorithm to every DP of the 4D-STEM tilt series  - giving a list of every peak´s 3D coordinates and intensities. The serial processing of the $3D_{dreal}$-$3D_{diff}$ reconstruction took almost 2 months to calculate on a normal laptop computer (2.2 Ghz, 3 cores) and resulted in almost 4 terabytes of data, whereas the $3D_{dreal}$-$3D_{peaks}$ took only a few days with a size of merely 1 Gb. These calculations could be rewritten for GPU for enhanced speed. Method 2 can be seen as a "sparse" version of Method 1.

Both of these methods rely on the back-projection scheme, where as a guide – an inverse Radon index map (nearest neighbor) was calculated, wherein the intensity value corresponds to the x-axis index of the 4D-STEM tilt image with appropriate tilt angle, and the y-axis refers to the 2D plane of the $3D_{dreal}$ z-axis (xy plane in Figure 6(a)). This map has the same dimensions as the $3D_{dreal}$ xy-plane and as is depicted in Figure 6(b). For every 4D-STEM tilt image the 2D DPs at specific x index are projected and accumulated to every voxel where the x-axis index and the value in the Radon map coincides along the (red) ray. Accumulating all 2D DPs on every x-axis indexes and tilt angles, then masking the results by multiplication with the bitmap segmentation of the $3D_{dreal}$ on that z-index yields an image $3D_{diff}$ as shown in Figure 6(b) on every voxel in the navigation space within the segmented $3D_{dreal}$. This process is repeated for every z-axis indexes.

\begin{figure*}[ht]
	\centering
	  \includegraphics[width=1.0\linewidth]{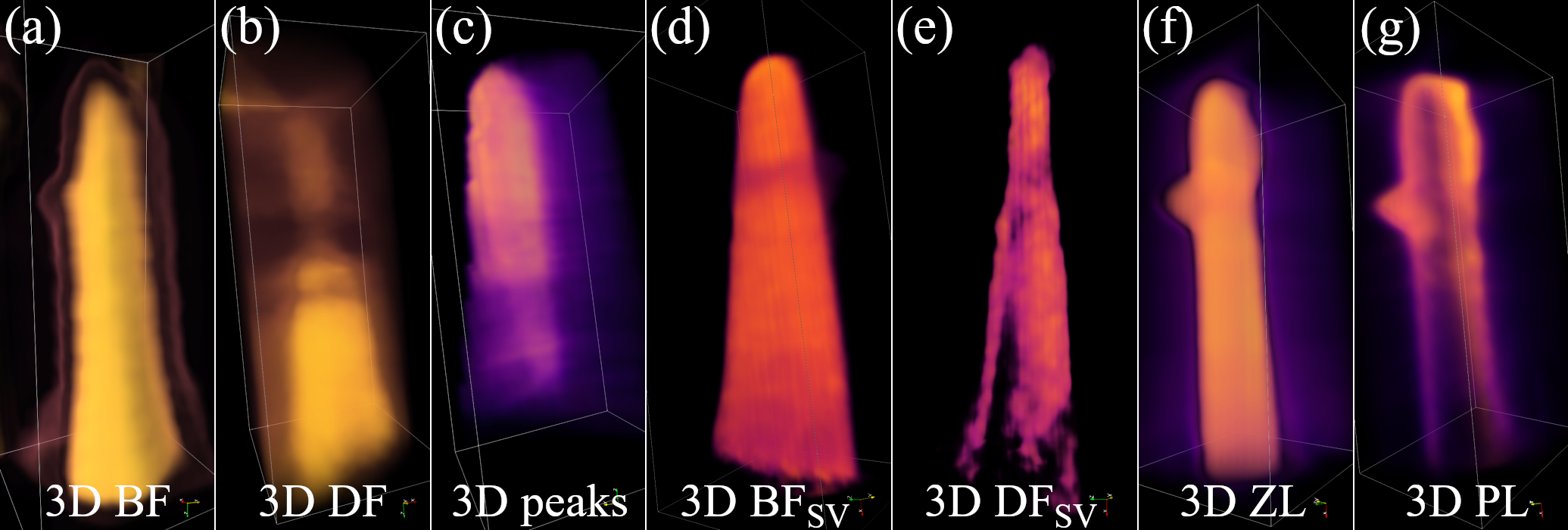}
	\caption{7D-STEM virtual reconstructions. (a) A bright field and (b) a dark field image taken from $3D_{diff}$ images using 3D spherical and annular masks respectively (method 1). (c) Sum of peaks image using method 2. (d) and (e) are reconstructed in TomViz using the single value virtual 4D-STEM tilt series while using a 2D spherical and annular masks respectively. (f) and (g) are similarly TomViz reconstructions using the sum of zero loss and plasmon peaks respectively.}
\end{figure*}

These 6D-STEM and 3D Peak object data can be readily used for 3D virtual imaging by either using a 3D spherical bitmap mask, or by just choosing specific peaks from the list to accumulate the intensities within the 3D virtual image respectively. However, this FIB milled TiNi sample was not diffracting enough to form meaningful dark field virtual images (Figure 7(b)), coupled with the back lash of reduced resolution when using basic back-projection and nearest neighbor interpolation scheme. The 3D virtual images in Figure 7 (a) and (b) were calculated by going through all the $3D_{diff}$ voxels of the $3D_{dreal}$ reconstruction, multiplying them with a spherical (BF) and annular mask (DF), and then calculating the sum. The Sum of Peaks reconstruction in Figure 7(c) was calculated similarly, except the sum of all detected peaks were used to formulate the $3D_{dreal}$ reconstruction.

A third possibility (Method 3) to create the reconstruction – and to make use of the more sophisticated single intensity value reconstruction techniques – is to first make 2D virtual images from the 4D-STEM data cubes, and construct a virtual image tilt series. In Figure 7 (d) and (e), a 3D virtual BF $3D$ $BF_{SV}$ and DF $3D$ $DF_{SV}$ image is shown, reconstructed from a virtual tilt series using TomViz. A 2D mask of 0–0.1 $\tAA^{-1}$ and 0.1–2.4 $\tAA^{-1}$ was used to calculate the sum on DP intensities within those annular ranges respectively. 

After the 4D-STEM acquisition at every tilt angle, an EELS SI was also acquired automatically using the St4DeM software. Following identical registration procedures as above, a full 3D reconstruction for every energy value could be then accomplished. Having an extra energy dimension makes this all together a 7-dimensional data set. However here we only used the sum of the zero-loss peak and plasmon peak respectively for reconstructions, which are given in Figure 7 (f) and (g) respectively.

The acquisition process of a 4D-STEM tilt series can take several hours, depending on the chosen tilt step increment (here 2 degrees). Hence one of the drawbacks of this method, is the uncertainty in the occurrence of sufficient and relevant diffraction peaks on each tilt step. In 3D EM this uncertainty is minimized by taking hundreds of fast images with a minimal tilt increment (usually 0.1 degrees) or by using continuous rotation scheme. A dedicated 4D-STEM $100k \fps$ direct electron detector would be ideal to mimic the amount of images taken using 3D EM assuring enough diffraction space information. Another option to circumvent this problem, would be to create on-line DP analysis and simulation software to predict the orientation of the crystal(s) and use a double tilt holder to tilt the sample into appropriate zone axes. This option would limit the range of tilt angles but would guarantee relevant and sufficient diffraction information.

\bigskip
\section{Conclusions}

A suite for 4D-STEM acquisition techniques has been created. Using software synchronization, acquisition speed of $100 \fps$ was demonstrated. This software also allows EFTEM zero-loss imaging, EDS and dual-EELS spectrum imaging. Additionally, useful 4D alignment and visualization methods were implemented including iDPC analysis. The software also allows the acquisition of 4D-STEM tomographic tilt series, and a 7D-STEM reconstruction pipeline was given as a proof of principle. 

\begin{acknowledgments}
The research leading to these results has received funding from
Swedish Foundation for Strategic Research (ITM17-0301).

\end{acknowledgments}
\bibliography{ref}
\end{document}